\documentclass[twocolumn,aps,pra,showpacs]{revtex4}
\usepackage{epsfig}
\begin{document}
\title{Improving teleportation of continuous variables by local operations}
\author{Ladislav Mi\v{s}ta Jr. and Radim Filip}
\affiliation{Department of Optics, Palack\' y University,\\
17. listopadu 50,  772~07 Olomouc, \\ Czech Republic}
\date{\today}
\begin{abstract}
We study a continuous-variable (CV) teleportation protocol based on a
shared entangled state produced by the quantum-nondemolition (QND)
interaction of two vacuum states. The scheme utilizes the
QND interaction or an unbalanced beam splitter in 
the Bell measurement. It is shown that in 
the non-unity gain regime the signal transfer coefficient can be enhanced 
while the conditional variance product remains preserved by applying appropriate 
local squeezing operation on sender's part of the shared entangled state. 
In the unity gain regime it is demonstrated that the fidelity of
teleportation can be increased with the help of the local squeezing 
operations on parts of the shared entangled state that convert effectively our scheme 
to the standard CV teleportation scheme. Further, it is proved
analytically that such a choice of the local symplectic operations 
minimizes the noise by which the mean number of photons in the input state is 
increased during the teleportation. Finally, our analysis reveals that the local 
symplectic operation on sender's side can be integrated into the Bell measurement 
if the interaction constant of the interaction in the Bell measurement can 
be adjusted properly. 

\end{abstract}

\maketitle

Quantum entanglement can be used for information carried by quantum 
objects to be processed in a unique way. For instance, entanglement enables 
a disembodied transfer of an unknown state of one quantum system to another 
system with precision that cannot be achieved using only classical resources. 
This the so called quantum teleportation was first proposed theoretically 
\cite{Bennett_93} and realized experimentally \cite{Bouwmeester_97} in the 
context of observables with discrete spectra. The studies of quantum 
teleportation did not restrict to the discrete variables but they were also 
extended by Vaidman \cite{Vaidman_94} into the realm of physical quantities 
with continuous spectra--continuous variables. A feasible implementation of 
Vaidman's continuous-variable (CV) teleportation protocol 
was then proposed by Braunstein and Kimble (BK) \cite{Braunstein_98}. 

In the BK protocol the role of the CVs is played by the canonically conjugate variables 
$x_{\rm in}$ and $p_{\rm in}$ of the input CV system ``in''
and the state to be teleported is an unknown coherent state. 
At the beginning of the BK protocol a sender Alice ($A$) and a receiver Bob ($B$) 
share a CV entangled state of two CV systems $A$ and $B$ described by the 
canonically conjugate variables $x_{A}$, $p_{A}$, $x_{B}$ and $p_{B}$ 
($[x_{i},p_{j}]=i\delta_{ij}$). At the first stage, Alice performs the 
so called Bell measurement on the systems $A$ and ``in'' by
superimposing them on a balanced beam splitter (BS) and detecting at its
outputs the variables $x_{\rm in}'=(x_{\rm in}+x_{A})/\sqrt{2}$ 
and $p_{A}'=(p_{\rm in}-p_{A})/\sqrt{2}$. She obtains certain classical values 
$\bar x_{\rm in}$ and $\bar p_{A}$ and sends them via classical channel to 
Bob. In order to recover the input variables $x_{\rm in}$ and $p_{\rm in}$ 
in his system Bob amplifies the measurement results by the gain factor $\sqrt{2}$ 
and performs the displacements $x_{B}\rightarrow x_{\rm out}=x_{B}+\sqrt{2}\bar x_{\rm in}$ 
and $p_{B}\rightarrow p_{\rm out}=p_{B}+\sqrt{2}\bar p_{A}$.
As a result, Bob's conjugate variables read as $x_{\rm out}=x_{\rm in}+(x_{A}+x_{B})$ and 
$p_{\rm out}=p_{\rm in}-(p_{A}-p_{B})$ \cite{Loock_00a} and thus the input variables 
were teleported to Bob with some added noises $x_{A}+x_{B}$ and $p_{A}-p_{B}$. 
Without entanglement Alice and Bob can achieve only 
limited quality of the teleportation. In the best case when the systems $A$ and $B$ 
are prepared in the vacuum states \cite{Braunstein_00} two vacuum units of noise are 
added into each of the input variables $x_{\rm in}$ and $p_{\rm in}$. This noise limits the quality of 
the teleportation to the classical regime. However, the noise can be reduced and therefore 
the so called quantum regime of teleportation can be achieved if Alice and Bob use a specific entangled state 
that possesses quantum correlations of equal strength between canonically conjugate variables
whose strength increases with increasing entanglement, i. e. 
$\langle(x_{A}+x_{B})^{2}\rangle=\langle(p_{A}-p_{B})^{2}\rangle=e^{-2\kappa}\rightarrow 0$ 
for $\kappa\rightarrow \infty$ ($\kappa>0$ is a squeezing parameter).
Obviously, sharing this type of entanglement Alice can teleport the input state to 
Bob with arbitrarily high precision if the entanglement is sufficiently strong.
This type of entanglement is called Einstein-Podolsky-Rosen (EPR) entanglement \cite{Einstein_35} 
and can be prepared by mixing the systems $A$ and $B$ equally squeezed in conjugate variables 
$x_{A}$ and $p_{B}$, respectively, on a balanced BS \cite{Walls_94}. 

To date, all the ingredients of the BK scheme encompassing beam splitters, squeezers and detectors of 
canonically conjugate variables were well managed only for light. Therefore, it was possible to demonstrate 
to date only CV quantum teleportation of light \cite{Furusawa_98}. In these experiments, the role of 
the CV systems was played by the single modes of an optical field and the CVs were realized by the
quadrature amplitudes of the field. Recently, however, a great attention has been paid to the teleportation
protocols for CVs of material objects \cite{Duan_00,Kuzmich_00}. These protocols involve two kinds 
of CV systems: coherent linearly polarized light pulses and spin-polarized macroscopic atomic samples. 
The respective CVs $x_{A}$ and $p_{A}$ are for a light pulse $A$ realized by the properly 
normalized components of the operator of the Stokes vector; for an atomic sample $B$ 
the respective CVs $x_{B}$ and $p_{B}$ are realized by the properly normalized components 
of the collective spin operator of the atomic sample \cite{Julsgaard_01}. Similarly as in the 
BK scheme, the protocols exploit the EPR entanglement of two atomic samples \cite{Duan_00} or 
the EPR entanglement of two light beams \cite{Kuzmich_00}. However, the specific feature 
of the two systems involved in these protocols is that they interact naturally via the 
quantum-nondemolition (QND) interaction \cite{Kuzmich_00} described by the interaction Hamiltonian
\begin{equation}\label{HQND}
H_{\rm QND}=-\kappa x_{A}p_{B},  
\end{equation}
where $\kappa$ is the coupling constant. As the BS interaction also 
the QND interaction of two orthogonally squeezed vacuum states produces the 
EPR entanglement \cite{Milburn_99}. More interestingly, the QND interaction produces 
the entangled state even from two vacuum states but this is no more the EPR state. 
Therefore, the use of such the QND entangled state in the BK protocol will not lead 
to its optimal performance. For this reason Horoshko and Kilin (HK) \cite{Horoshko_00}
studied the teleportation protocol based on sharing of the QND entanglement and 
utilizing an unbalanced BS in the Bell measurement. They have shown that quantum 
regime of teleportation can be achieved if sufficiently strong QND entanglement is 
available.  
 
In this article we investigate a generalized CV teleportation protocol that contains the
HK protocol as a particular case. Our protocol is based on the QND entanglement 
and utilizes either the QND interaction or an unbalanced BS in the 
Bell measurement. We study both the non-unity gain and the unity gain regimes. 
The non-unity gain regime is characterized by the conditional variance product $V$ 
and the signal transfer coefficient $T$ \cite{Ralph_98}.  
It is shown on two particular examples that for a fixed interaction 
in the Bell measurement the parameter $T$ can be enhanced while preserving 
the parameter $V$ and thus the performance of our protocol can be improved  
by applying a suitable squeezing operation on Alice's part of the shared entangled state. 
The unity gain regime is characterize by the fidelity $F$. In contrast with the HK 
scheme our scheme allows to achieve quantum regime of teleportation when $F>1/2$ for 
arbitrarily small amount of the shared QND entanglement. For a particular amount of the shared 
entanglement corresponding to the feasible interaction constant $\kappa t=1$ \cite{Fiurasek_04}
($t$ is the interaction time) we show analytically that the use of a 
suitable unbalanced BS in the Bell measurement provides a higher 
teleportation fidelity than the use of a balanced 
BS as proposed by HK. Recently, it was 
demonstrated in \cite{Bowen_03b} that in the teleportation protocol exploiting 
entanglement produced from a single squeezed state and a balanced BS in the Bell 
measurement one can increase the teleportation fidelity by applying local squeezers 
on parts of the shared entangled state. Here we show that the fidelity in our scheme can 
be increased by application of local squeezing operations on both parts of the 
shared entangled state that convert effectively the scheme to the BK scheme 
\cite{Braunstein_98}. Moreover, it is proved analytically that if we characterize the unity
gain teleportation by the noise ${\cal N}$ by which the mean number of photons in the input state 
is increased during the teleportation then such a choice of the local operations is optimal.  

The paper is organized as follows. In Section~\ref{sec_QND} we study a teleportation 
scheme utilizing the QND entanglement as an entanglement resource and
the QND interaction or an unbalanced BS in the Bell measurement. Section~\ref{sec_generic} 
deals with a generic CV teleportation scheme and the optimization of the noise 
${\cal N}$ with respect to the local symplectic transformations is performed. 
Section~\ref{sec_conclusion} contains conclusion.
\begin{figure}
\centerline{\psfig{width=8.0cm,angle=0,file=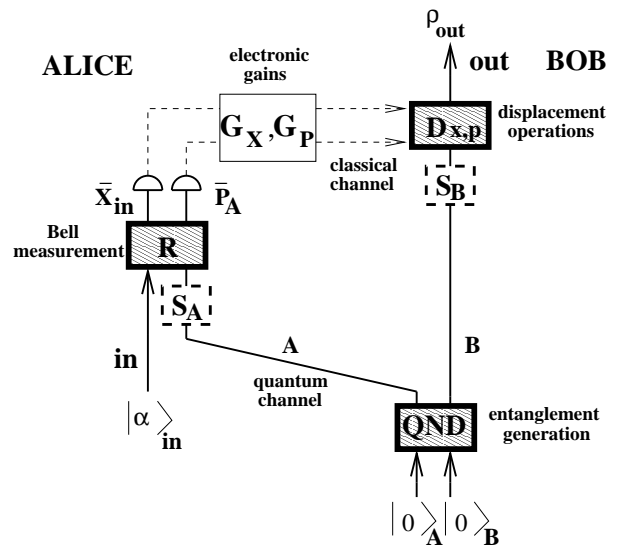}}
\caption{Schematic of the CV telportation using the
QND interaction. A bipartite entangled state 
produced by the QND interaction (\ref{HQND}) of two vacuum states
$|0\rangle_{A}$ and $|0\rangle_{B}$ is shared by Alice and Bob. $|\alpha\rangle_{\rm in}$, 
an unknown input coherent state; $\rho_{\rm out}$, output state; 
$S_{A}$ and $S_{B}$, auxiliary squeezing operations allowing to improve the quality of teleportation; 
$R$, the QND interaction or an unbalanced BS; $G_{x}$, $G_{p}$, electronic gains for the transformation 
from the classical measurement results $\bar x_{\rm in}$ and $\bar p_{A}$ to Bob's output system.}
\label{fig1}
\end{figure}
\section{Quantum teleportation with quantum-nondemolition interaction}\label{sec_QND}

The scheme of our teleportation protocol is depicted in Fig.~\ref{fig1}.
If not stated explicitly otherwise an optical terminology is used throughout 
the article in which the CV systems involved are single modes of an 
optical field and the role of the canonically conjugate variables is played by 
the quadrature amplitudes of the optical field. We use the Heisenberg picture 
description of teleportation as introduced in \cite{Loock_00a}.

At the beginning of our protocol, Alice and Bob share the state of two modes $A$ and $B$ 
produced by the QND interaction (\ref{HQND}) of two vacuum states $|0\rangle_{A}$ and $|0\rangle_{B}$. 
In Heisenberg picture, the quadratures $x_{A}^{(0)}$, $p_{A}^{(0)}$, $x_{B}^{(0)}$ and 
$p_{B}^{(0)}$ of the two vacuum modes transform as
\begin{eqnarray}\label{QNDHeis}
x_{A}=x_{A}^{(0)}, \quad p_{A}=p_{A}^{(0)}+gp_{B}^{(0)},\nonumber\\
x_{B}=x_{B}^{(0)}-gx_{A}^{(0)},\quad p_{B}=p_{B}^{(0)},  
\end{eqnarray}
where $g=\kappa t$ ($t$ is the interaction time) is the interaction constant. 
For any $g>0$ the produced state is an entangled state whose entanglement increases with increasing 
$g$. This can be seen by noting that the purity of the reduced state of the mode $A$ 
given by the formula $P_{A}=1/(2\sqrt{\langle x_{A}^{2}\rangle\langle p_{A}^{2}\rangle})=
1/\sqrt{1+g^{2}}$ is less than unity for any $g>0$ and goes to zero with increasing 
$g$. Here and in the rest of this section the angle brackets denote the averaging over 
the initial state $|\alpha\rangle_{\rm in}|0\rangle_{A}|0\rangle_{B}$.
The produced entanglement is not the EPR entanglement as 
$\langle(x_{A}+x_{B})^{2}\rangle=\langle(p_{A}-p_{B})^{2}\rangle=[(g-1)^2+1]/2\geq 1/2$
and therefore it does not ``match'' the BK scheme. For this reason, Alice performs the 
Bell measurement that differs from the Bell measurement used in the BK scheme. 
She lets her mode $A$ to interact with the input mode ``in'' whose unknown coherent state
$|\alpha \rangle_{\rm in}$ is to be teleported in another QND interaction 
$H_{\rm QND}'=\kappa' x_{A}p_{\rm in}$ and subsequently measures the 
quadratures $x_{\rm in}'=x_{\rm in}+g'x_{A}$ and $p_{A}'=p_{A}-g'p_{\rm in}$ 
($g'=\kappa' t'$). The quadratures $x_{A}$ and $p_{\rm in}$ are the so-called QND variables 
that are not affected by the interaction $H_{\rm QND}'$. This interaction, however, 
establishes correlations between $p_{A}$ and $p_{\rm in}$. If the mode $A$ were 
infinitely squeezed in the quadrature $p_{A}$ the measurement of $p_{A}'$ 
would provide result proportional to the result of the measurement of $p_{\rm in}$. 
This is the principle of the QND measurement. Thus, the Bell measurement performed 
by Alice in fact consists of the QND measurement of the quadrature $p_{\rm in}$ with 
the ancillary mode $A$ being a part of an entangled system and the
direct measurement of the quadrature $x_{\rm in}$ at the output of the
QND interaction. After the measurement Alice has certain classical values 
$\bar x_{\rm in}$ and $\bar p_{A}$ and sends them via classical channel to Bob 
who performs the displacements (denoted $D_{x,p}$ in Fig.~\ref{fig1}) 
$x_{B}\rightarrow x_{\rm out}=x_{B}+G_{x}\bar x_{\rm in}$ and 
$p_{B}\rightarrow p_{\rm out}=p_{B}-(G_{p}/g')\bar p_{A}$, where
the parameters $G_{x}$ and $G_{p}$ describe normalized gains.
As a result, Bob's output quadratures read as  
\begin{eqnarray}\label{output}
x_{\rm out}=G_{x}x_{\rm in}+X,\quad p_{\rm out}=G_{p}p_{\rm in}-P, 
\end{eqnarray}
where
\begin{eqnarray}\label{XP}
X&=&G_{x}g'x_{A}+x_{B}=\left(G_{x}g'-g\right)x_{A}^{(0)}+x_{B}^{(0)},\nonumber\\
P&=&\frac{G_{p}}{g'}p_{A}-p_{B}=\frac{G_{p}}{g'}p_{A}^{(0)}-\left(1-G_{p}\frac{g}{g'}\right)p_{B}^{(0)},
\end{eqnarray}
are the quadrature added noises. Note, that instead of using the QND interaction in the Bell measurement Alice 
can use equally well an unbalanced $BS$ with reflectivity ${\cal R}$ and transmissivity ${\cal T}$. 
In this case, she detects the quadratures $x_{\rm in}'={\cal R}x_{A}+{\cal T}x_{\rm in}$, 
$p_{A}'=-{\cal R}p_{\rm in}+{\cal T}p_{A}$ and Bob performs the displacements 
$x_{B}\rightarrow x_{\rm out}=x_{B}+(G_{x}/{\cal T})\bar x_{\rm in}$ and 
$p_{B}\rightarrow p_{\rm out}=p_{B}-(G_{p}/{\cal R})\bar p_{A}$. Consequently, Bob's output 
quadratures are given by the Eqs.~(\ref{output}) and (\ref{XP}), where $g'$ is replaced by 
${\cal R}/{\cal T}$.

In order to quantitatively asses the quality of our teleportation 
it is convenient to distinguish two different regimes of teleportation.
\subsection{Non-unity gain teleportation}

The quality of teleportation is often expressed quantitatively by the
fidelity $F=\!\!_{\rm in}\!\langle\alpha|\rho_{\rm out}|\alpha\rangle_{\rm in}$,
where $\rho_{\rm out}$ is the output state. In the non-unity gain regimes
($G_{x}\ne 1, G_{p}\ne 1$), however, the fidelity strongly depends on gains 
$G_{x}$ and $G_{p}$ and usually it decreases very quickly as increases the deviation 
of the gains from unity \cite{Bowen_03}. Therefore, there can exist non-unity gain 
regimes having quantum nature that are not captured by the fidelity. 
For this reason it was introduced a more detailed characterization of 
teleportation with the help of the quantities conditional variance 
product $V$ and the signal transfer coefficient $T$ \cite{Ralph_98,Bowen_03}. 
The parameter $V$ is defined by the formula 
\begin{eqnarray}\label{V}
V=V_{\rm out|in}^{x}V_{\rm out|in}^{p},
\end{eqnarray}
where $V_{\rm out|in}^{x}=\langle\left(\Delta x_{\rm out}\right)^{2}\rangle
-G_{x}^{2}\langle\left(\Delta x_{\rm in}\right)^{2}\rangle$ and  
$V_{\rm out|in}^{p}=\langle\left(\Delta p_{\rm out}\right)^{2}\rangle
-G_{p}^{2}\langle\left(\Delta p_{\rm in}\right)^{2}\rangle$ are the 
quadrature conditional variances between the input and output states.
The conditional variance product $V$ describes the noise added into 
the input state during the teleportation process. Without entanglement 
the noise is limited from below by the fluctuations in Bob's mode and 
therefore $V\geq1/4$ \cite{Bowen_03}. $V=0$ indicates teleportation with no added noise. 

The signal transfer coefficient is defined as a sum 
$T=T_{x}+T_{p}$ of the quadrature signal transfer coefficients 
$T_{i}=SNR_{\rm out}^{i}/SNR_{\rm in}^{i}$ ($i=x,p$), where $SNR^{i}$
denotes conventional signal-to-noise ratio for quadrature $i$.  
For teleportation of coherent states the parameter $T$ can be expressed 
in the form \cite{Poizat_94}
\begin{eqnarray}\label{T}
T=\frac{G_{x}^{2}}{G_{x}^{2}+2V_{\rm out|in}^{x}}+\frac{G_{p}^{2}}{G_{p}^{2}+2V_{\rm out|in}^{p}}.
\end{eqnarray}
In the absence of entanglement the intrinsic fluctuations of Alice's
mode limit the transfer coefficient to $T\leq 1$ \cite{Bowen_03}.  
On the other hand $T=2$ corresponds to a perfect signal transfer. 
With respect to the properties of the parameters $T$ and $V$ it is thus natural to 
define quantum regime of teleportation as a regime in which $V<1/4$ and simultaneously 
$T>1$.

In our protocol the added noises (\ref{XP}) are uncorrelated with 
the input state. Hence, $V_{\rm out|in}^{x}=\langle X^{2}\rangle$, 
$V_{\rm out|in}^{p}=\langle P^{2}\rangle$ and therefore
\begin{eqnarray}
V&=&\langle X^{2}\rangle\langle P^{2}\rangle,\label{Vour}\\
T&=&\frac{G_{x}^{2}}{G_{x}^{2}+2\langle X^{2}\rangle}+
\frac{G_{p}^{2}}{G_{p}^{2}+2\langle P^{2}\rangle}\label{Tour},
\end{eqnarray}
where 
\begin{eqnarray}\label{added}
\langle X^{2}\rangle&=&\left[(G_{x}g'-g)^{2}+1\right]/2,\nonumber\\ 
\langle P^{2}\rangle&=&\left[(G_{p}/g')^{2}+(1-G_{p}g/g')^{2}\right]/2 
\end{eqnarray}
are variances calculated from the Eq.~(\ref{XP}). 

In what follows we restrict our attention to two particular non-unity
gain regimes. For these two cases we then show among other things 
that if our scheme operates in quantum regime when $V<1/4$ and $T>1$ then 
the signal transfer coefficient $T$ can be increased while preserving the 
conditional variance product $V$ by application of a suitable local 
squeezing operation on Alice's mode.

1. The regime when the output state has minimum additional noise, i.
e. the gains $G_{x}$ and $G_{p}$ are adjusted in such a way that 
$V=V_{\rm min}\equiv\mbox{min}_{G_{x},G_{p}}V$. Deriving the
Eq.~(\ref{Vour}) with respect to $G_{x}$ and $G_{p}$, setting the
obtained expressions equal to zero and solving these equations one finds
that the parameter $V$ is minimized by $G_{x,{\rm min}}=g/g'$ and 
$G_{p,{\rm min}}=gg'/(1+g^2)$. For such gains the parameters $V$ and $T$ read as
\begin{eqnarray}
V_{\rm min}&=&\frac{1}{4(1+g^{2})},\label{Vmin}\\ 
T_{\rm V_{min}}&=&1+\frac{g'^{2}\left(g^{2}-\frac{\sqrt{5}+1}{2}\right)
\left(g^{2}+\frac{\sqrt{5}-1}{2}\right)}{(g^{2}+g'^{2})(1+g^{2}+g^{2}g'^{2})}\label{Tmin}. 
\end{eqnarray}
Apparently, since $V_{\rm min}<1/4$ for any $g>0$ our scheme operates 
from the point of view of the parameter $V$ in quantum regime as soon
as there is a nonzero shared entanglement. However, the
Eq.~(\ref{Tmin}) reveals that our scheme in general does not operate in the quantum 
regime from the point of view of the parameter $T$ as $T_{\rm V_{min}}>1$ 
only if $g>\sqrt{(\sqrt{5}+1)/2}\approx 1.27$.  
Hence, in order to fulfill both the conditions $V_{\rm min}<1/4$ and 
$T_{\rm V_{min}}>1$ simultaneously and thus to reach in this particular case the 
quantum regime of teleportation the shared entanglement must be sufficiently large. The parameter 
$T_{\rm V_{min}}$ depends on the asymmetry parameter $g'$ of the Bell measurement 
and can be maximized with respect to it. Solving the extremal equation 
$dT_{\rm V_{min}}/dg'=0$ one finds that $T_{\rm V_{min}}$ has a maximum if 
$g>\sqrt{(\sqrt{5}+1)/2}$ and it is localized in the point 
$g'_{\rm opt}=(1+g^2)^{\frac{1}{4}}$. Substituting $g'_{\rm opt}$ to 
the Eq.~(\ref{Tmin}) we arrive at the optimal signal transfer coefficient     
\begin{eqnarray}\label{Tminopt}
T_{\rm V_{min},opt}=\frac{2g^2}{g^{2}+\sqrt{1+g^{2}}}. 
\end{eqnarray}
The latter analysis indicates that if $g'\ne g'_{\rm opt}$ the parameter 
$T_{\rm V_{min}}$ can be increased to the optimal value (\ref{Tminopt}) 
by adjusting $g'$ in the Bell measurement to the optimal value $g'_{\rm opt}$.
If the Bell measurement is fixed, i. e. $g'$ is fixed, the optimal
value $g'_{\rm opt}$ can be adjusted by a suitable squeezing operation
on Alice's mode. Namely, the squeezing operation $x_{A}\rightarrow e^{r_{A}}x_{A}$ 
and $p_{A}\rightarrow e^{-r_{A}}p_{A}$ ($r_{A}$ is the squeezing parameter) 
transforms effectively $g'$ in 
Eq.~(\ref{XP}) to $\tilde g=e^{r_{A}}g'$. Therefore, if the squeezing operation is
such that $e^{r_{A}}=(1+g^2)^{\frac{1}{4}}/g'$ we attain optimal
value of the signal transfer coefficient (\ref{Tminopt}).       
As the parameter $V_{\rm min}$ remains preserved when changing $g'$ 
(it does not depend on $g'$) we get deeper into the quantum region of 
teleportation and thus improve the quality of teleportation in our
protocol. For example, if we take $g=2.5$ and the Bell measurement is
realized by the balanced $BS$ for which $g'={\cal R}/{\cal T}=1$ the 
formula (\ref{Tmin}) gives $T_{\rm V_{min}}\approx 1.32$ whereas 
for the unbalanced BS with $g'_{\rm opt}=(1+g^{2})^{\frac{1}{4}}\approx 
1.64$ it gives a higher signal transfer coefficient $T_{\rm V_{min},opt}\approx 1.4$. 
In addition, for the scheme with $g'=1$ $V_{\rm min}\rightarrow
0$ but $T_{\rm V_{min},opt}\rightarrow 1+g'^{2}/(1+g'^{2})=1.5$ 
with increasing $g$ in contrast with the improved scheme where 
$V_{\rm min}\rightarrow0$ and $T_{\rm V_{min},opt}\rightarrow 2$ with 
increasing $g$ and therefore we approach perfect teleportation from 
the point of view of the parameters $T$ and $V$ with increasing 
shared entanglement. 

2. The regime when the input signals are transferred optimally,  
i. e. the gains $G_{x}$ and $G_{p}$ are adjusted in such a way that 
$T=T_{\rm max}\equiv\mbox{max}_{G_{x},G_{p}}T$. Solving the extremal
equations $\partial T/\partial G_{i}=0$ ($i=x,p$), where $T$ is given in Eq.~(\ref{Tour}) 
one finds that the parameter $T$ is maximized by $G_{x,{\rm max}}=(1+g^2)/gg'$ and 
$G_{p,{\rm max}}=g'/g$. For such gains the parameters $T$ and $V$ read as
\begin{eqnarray}
T_{\rm max}&=&1+\frac{g^{2}g'^{2}}{(1+g'^{2})(1+g^{2}+g'^{2})},\label{Tmax}\\
V_{\rm T_{max}}&=&\frac{1}{4}\left(\frac{1}{g^{2}}+\frac{1}{g^{4}}\right)\label{Vmax}. 
\end{eqnarray}
Clearly, $T_{\rm max}>1$ for any $g,g'>0$ and therefore our scheme
operates
in the quantum regime of teleportation from the point of view of the
parameter $T$ once there is some shared entanglement. On the other
hand, the Eq.~(\ref{Vmax}) reveals that $V_{\rm T_{max}}<1/4$ only if
$g^{4}-g^{2}-1=[g^{2}-(\sqrt{5}+1)/2][g^{2}+(\sqrt{5}-1)/2]>0$.
Hence, like in the previous case our scheme works in the quantum regime of
teleportation when $T_{\rm max}>1$ and simultaneously $V_{\rm T_{max}}<1/4$ 
only if $g>\sqrt{(\sqrt{5}+1)/2}\approx1.27$, i. e. only if
the shared entanglement is strong enough. Further, as in the 
previous case we can maximize $T_{\rm max}$ with respect to $g'$.
Solving the respective extremal equation $dT_{\rm max}/dg'=0$ one finds 
that $T_{\rm max}$ attains maximum at the same point 
$g'_{\rm opt}=(1+g^2)^{\frac{1}{4}}$ as $T_{\rm min}$ and it is equal to
\begin{eqnarray}\label{Tmaxopt}
T_{\rm max,opt}=\frac{2\sqrt{1+g^2}}{1+\sqrt{1+g^2}}. 
\end{eqnarray}
Similarly as in the case 1 this optimal value of the signal transfer 
coefficient can be reached by adjusting $g'$ in the Bell measurement to
the $g'_{\rm opt}$ or, for fixed $g'$ in the Bell measurement, by using 
the same squeezing operation on Alice's mode as in the case 1.
Further, for $g=2.5$ and for the balanced BS in the Bell measurement ($g'=1$) the 
Eq.~(\ref{Tmax}) leads to the signal transfer coefficient $T_{\rm max}\approx1.38$ 
while for the unbalanced BS with $g'_{\rm opt}\approx 1.64$ 
we obtain using the Eq.~(\ref{Tmaxopt}) a higher value $T_{\rm max,opt}\approx1.46$.  
Finally, in both the particular cases $V_{\rm T_{max}}\rightarrow 0$ 
with increasing $g$ whereas $T_{\rm max}\rightarrow 1+g'^{2}/(1+g'^{2})=1.5$ and
$T_{\rm max,opt}\rightarrow 2$ and therefore only the improved scheme 
approaches perfect teleportation from the point of view of the parameters 
$T$ and $V$ with increasing shared entanglement. 

\subsection{Unity gain teleportation}

In this regime the gains are adjusted in such a way that $G_{x}=G_{p}=1$. 
Then, it follows from the Eqs.~(\ref{output}) and (\ref{XP}) that
the first moments of the input state are preserved, i. e. 
$\langle x_{\rm out}\rangle=\langle x_{\rm in}\rangle$ and 
$\langle p_{\rm out}\rangle=\langle p_{\rm in}\rangle$. 
This implies, that in the unity gain regime the quality of our teleportation 
depends only on the added noises (\ref{XP}) and therefore it can be conveniently described by 
the fidelity $F=\!\!_{\rm in}\!\langle\alpha|\rho_{\rm out}|\alpha\rangle_{\rm in}$. 
$F=1$ corresponds to a perfect teleportation while 
$F>1/2$ indicates quantum regime of teleportation \cite{Braunstein_00}. Since in our protocol the 
added noises (\ref{XP}) are uncorrelated both mutually and also with the input state the fidelity 
can be expressed as \cite{Bowen_03}
\begin{eqnarray}\label{FQND1}
F=\frac{1}{\sqrt{\left(1+\langle X^{2}\rangle\right)\left(1+\langle P^{2}\rangle\right)}}.
\end{eqnarray} 
Calculating the variances $\langle X^{2}\rangle$ and $\langle P^{2}\rangle$  using the Eqs.~(\ref{added}) 
where $G_{x}=G_{p}=1$ we finally arrive at the fidelity of teleportation of our scheme in the form:
\begin{eqnarray}\label{F1}
F=\frac{2}{\sqrt{\left[2+(1/g')^{2}+(g/g'-1)^{2}\right]\left[3+(g-g')^{2}\right]}}.
\end{eqnarray}
Several important properties of our teleportation protocol can be derived from the latter formula.  
First, it reveals that if the parameter $g'$ in the Bell measurement can be 
adjusted properly then the quantum regime of teleportation can be attained for 
an arbitrarily small amount of the shared entanglement. This follows from the fact 
that if we put, e. g., $g'=g+1$ in Eq.~(\ref{F1}) then $F>1/2$
for any $g>0$. Second, as the denominator in the Eq.~(\ref{F1}) is always greater or 
equal to $\sqrt{2}\sqrt{3}$ the maximum achievable fidelity is bounded by the 
value $F_{\rm max}=\sqrt{2/3}\approx0.816$ that can be achieved, e. g., in the limit 
$g=g'\rightarrow\infty$. Finally, the formula (\ref{F1}) also reveals that 
although one could be tempted to think that for a given $g$ maximal fidelity  
is obtained for $g=g'$ this is not the case and the fidelity is maximized by $g'$ that 
in general {\it differs} from $g$. To illustrate this let us consider the particular 
value of the interaction constant $g=1$ that is well within reach of the current 
experiment \cite{Fiurasek_04}. Setting the derivative of the Eq.~(\ref{F1}) with respect to $g'$
equal to zero and solving the obtained extremal equation one finds that $F$ attains 
maximum for $g'=4/3>g=1$. On inserting $g'=4/3$ back into the Eq.~(\ref{F1}) for 
$g=1$ we finally get the maximum fidelity equal to $F_{1}=2\sqrt{6}/7\approx 0.7$.

Our results can be compared with the results obtained by Horoshko and Kilin (HK) 
\cite{Horoshko_00}. As in our protocol also in the HK protocol the shared entanglement 
is produced by the QND interaction (\ref{HQND}) and the Bell measurement utilizes an unbalanced BS.
In contrast with our protocol, in the HK protocol the asymmetry parameter ${\cal R}/{\cal T}$ 
of the BS and the interaction constant $g$ are tied together by the relation ${\cal R}/{\cal T}=g$.
Therefore, for $g=g'$ the formula (\ref{F1}) reduces to the fidelity of teleportation of 
coherent states in the HK protocol 
\begin{eqnarray}\label{FHK}
F_{\rm HK}=\frac{2}{\sqrt{3(2+1/g^{2})}}.
\end{eqnarray}
Clearly, like in our teleportation protocol the maximum teleportation fidelity is 
$F_{\rm HK,max}=\sqrt{2/3}$ and it is achieved in the limit $g\rightarrow\infty$. 
However, in contrast to our protocol, quantum teleportation can be achieved only 
if $g>\sqrt{3/10}\approx0.548$, i. e. $F_{\rm HK}>1/2$ only if sufficiently large shared 
entanglement is available. In addition, our results reveal that for $g=1$ our 
scheme provides a higher teleportation fidelity $F_{1}\approx 0.7$ than the HK scheme that
gives only $F_{{\rm HK},1}=2/3\approx 0.667$. Therefore, in the teleportation scheme using the 
shared entanglement produced by the QND interaction (\ref{HQND}) with $g=1$ of two vacua, 
a higher teleportation fidelity is obtained in the scheme with an unbalanced BS in the Bell 
measurement with ${\cal T}=3/5$ and ${\cal R}=4/5$ than in the HK scheme that exploits 
the balanced BS.

The previous results reveal that no matter what is the value of the parameter $g'$ 
in the Bell measurement and how strong is the shared entanglement the highest 
possible fidelity $F_{\rm max}=\sqrt{2/3}$ is always less than unity. In other words, 
the quality of transfer in our teleportation protocol is limited even in the limit of 
the infinitely large shared entanglement. This behaviour is due to the added noise 
$X$ given in Eq.~(\ref{XP}) that always contains at least one unit of the vacuum noise originating 
from the term $x_{B}^{(0)}$. However, this undesirable noise can be reduced and therefore 
the fidelity of our scheme can be increased if Alice and Bob transform the shared state by the local 
squeezing operations (denoted as $S_{A}$ and $S_{B}$ in Fig.~\ref{fig1}) $x_{A}\rightarrow (a/g')x_{A}$, 
$p_{A}\rightarrow (g'/a)p_{A}$, $x_{B}\rightarrow(1/a)x_{B}$ and $p_{B}\rightarrow a p_{B}$, where 
$a=(1+g^{2})^{\frac{1}{4}}$. These operations transform the original
added noises $X=g'x_{A}+x_{B}$ and $P=(1/g')p_{A}-p_{B}$ into the new added 
noises $X'=ax_{A}+(1/a)x_{B}$ and $P'=(1/a)p_{A}-ap_{B}$. Calculating
the variances $\langle X'^{2}\rangle$ and $\langle P'^{2}\rangle$ using the Eq.~(\ref{QNDHeis}) 
one finds that $\langle X'^{2}\rangle=\langle P'^{2}\rangle =\sqrt{1+g^{2}}-g\rightarrow 0$ 
with increasing $g$ and therefore the new added noises vanish with
increasing entanglement. Hence, by using the Eq.~(\ref{FQND1}) we finally arrive at the 
fidelity of teleportation in the improved scheme of the form:
\begin{eqnarray}\label{F2}
F_{S}=\frac{1}{1+\sqrt{1+g^{2}}-g}.
\end{eqnarray}
Obviously, the Eq.~(\ref{F2}) demonstrates that the quantum regime of teleportation when 
$F_{S}>1/2$ is again reached for any $g>0$. Moreover, since $\sqrt{1+g^{2}}-g\rightarrow 0$ 
with increasing $g$ $F\rightarrow 1$ with increasing $g$ and the output state converges to the 
input state with increasing entanglement similarly as in the BK scheme. In fact, the formula 
(\ref{F2}) can be rewritten in the form of the fidelity of the BK scheme 
$F_{\rm BK}=1/(1+e^{-2\kappa})$ \cite{Braunstein_01}, where 
$\kappa=(-1/2)\ln\left(\sqrt{1+g^{2}}-g\right)$ and thus the squeezing operations 
$S_{A}$ and $S_{B}$ were chosen such that they effectively transform our scheme 
into the BK scheme. Characterizing our teleportation protocol by the
noise by which the mean number of photons in the input state is
increased ${\cal N}=(\langle X^{2}\rangle+\langle P^{2}\rangle)/2$ 
instead of the fidelity (\ref{FQND1}) such a choice of $S_{A}$ and $S_{B}$ maximizes 
${\cal N}$ and therefore our improved scheme is optimal from the point of view of the 
parameter ${\cal N}$. The analytical proof of the latter statement is given in 
the next section.

We have seen that in some cases the application of suitable local squeezing 
operations on parts of the shared entangled state can improve considerably 
the efficiency of the CV teleportation. Despite of the fact that both the 
squeezed state of a light beam \cite{Slusher_85} and a squeezed state of an 
atomic sample \cite{Hald_99} were already realized experimentally it 
may be still an experimental challenge to implement the squeezing 
operations on parts of the entangled state. The obstacle can be partially 
avoided and Alice's squeezing operation $S_{A}$ can be saved if one can adjust 
properly the interaction constant $g'$ (or the ratio ${\cal R}/{\cal T}$) 
in the Bell measurement. To be more specific, if in our case we put 
$g'=a=(1+g^{2})^{\frac{1}{4}}$ (${\cal R}/{\cal T}=a=(1+g^{2})^{\frac{1}{4}}$)
the interaction in the Bell measurement effectively realizes the
needed squeezing operation $S_{A}$. 

Contrary to the operation $S_{A}$ the squeezing operation $S_{B}$ on Bob's side 
is in general inevitable, in particular if one needs to preserve the whole 
quantum state $\rho_{\rm out}$ for further processing. However, the squeezing
operation $S_{B}$ can be avoided if the input state $|\alpha\rangle_{\rm in}$ 
just carries some classical information encoded into its complementary 
quadratures $x_{\rm in}$ and $p_{\rm in}$ that is red from the output state $\rho_{\rm out}$ 
immediately after the teleportation by homodyne measurement of 
either of the output quadratures $x_{\rm out}$ or $p_{\rm out}$. 
Namely, formally the operation $S_{B}$ can be moved behind the displacement 
transformation $D_{x,p}$ (see Fig.~\ref{fig1}) that then must be changed to  
$x_{B}\rightarrow x_{\rm out}=x_{B}+a\bar x_{\rm in}$ and 
$p_{B}\rightarrow p_{\rm out}=p_{B}-1/(ag')\bar p_{A}$. The needed 
squeezing operation $S_{B}$ then can be realized merely as a scaling 
$\bar x_{\rm out}\rightarrow (1/a)\bar x_{\rm out}$ or 
$\bar p_{\rm out}\rightarrow a\bar p_{\rm out}$ of the 
classical outcomes $\bar x_{\rm out}$ and $\bar p_{\rm out}$ 
of the homodyne measurement.
\section{Generalized continuous-variable teleportation protocol}\label{sec_generic}

In this section we will prove analytically that the improvement of our
previous unity gain teleportation scheme by local squeezing operations that
effectively transform the scheme into the BK scheme is in a certain
sense that is specified below optimal. For this reason, let us consider 
the generalization of the protocol depicted in Fig.~\ref{fig1}. 
In the generalized scheme, Alice and Bob share a Gaussian state
of two modes $A$ and $B$ $\rho_{AB}$ with vanishing first moments, i.
e. $\langle\xi_k\rangle=\mbox{Tr}(\rho_{AB}\xi_k)=0$ ($k=1,\ldots,4$), where 
$\xi=(x_{A},p_{A},x_{B},p_{B})^{\rm T}$ is the column vector of the
quadratures $x_{A}$, $p_{A}$, $x_{B}$ and $p_{B}$. Such a state is completely 
characterized by the variance matrix $V_{AB}$ with elements 
$(V_{AB})_{kl}=\langle\{\Delta \xi_{k},\Delta\xi_{l}\}\rangle$, 
where $\Delta\xi_{k}=\xi_{k}-\langle\xi_{k}\rangle$ and $\{A,B\}\equiv(1/2)(AB+BA)$.
We assume that Alice and Bob can apply locally on their parts of the
shared state arbitrary single-mode linear transformations $S_{A}$ and $S_{B}$ of the quadratures 
$x_{i},p_{i}$ ($i=A,B$) preserving the canonical commutation rules $[x_{i},p_{j}]=i\delta_{ij}$. 
In matrix notation these transformations can be expressed by the formulas 
\begin{equation}\label{SA}
\xi_{A}'=S_{A}\xi_{A},\quad \xi_{B}'=S_{B}\xi_{B},    
\end{equation}
where $\xi_{i}=(x_{i},p_{i})^{\rm T}$,  $\xi_{i}'=(x_{i}',p_{i}')^{\rm T}$ ($i=A,B$) 
and $S_{i}$ are real $2\times 2$ matrices satisfying the so called
symplectic condition 
\begin{equation}\label{singlesympl}
S_{i}JS_{i}^{\rm T}=J,\quad J=
\left(\begin{array}{cc}
0 & 1 \\
-1 & 0 \\
\end{array} \right).
\end{equation}
These transformations involve single-mode squeezers and phase shifters as particular instances
and are conventionally denoted as symplectic transformations. Then, Alice performs the Bell 
measurement. She lets her mode $A$ and the mode ``in'' to interact 
in a two-mode interaction $R$ described by the quadratic Hamiltonian of the form 
$H_{R}=\sum_{i,j=1}^{4}a_{ij}\zeta_{i}\zeta_{j}$, where $a_{ij}$ are real coupling  
constants and $\zeta_{i}$ ($i=1,\ldots,4$) are components of the column vector 
$\zeta=(x_{A}',p_{A}',x_{\rm in},p_{\rm in})^{\rm T}$. In Heisenberg picture the interaction is 
described by the two-mode linear symplectic transformation of the form \cite{Simon_94}:
\begin{equation}\label{R}
\zeta'=R\zeta,\quad  
\end{equation}
where $\zeta'=(x_{A}'',p_{A}'',x_{\rm in}',p_{\rm in}')^{\rm T}$ and $R$ is a real 
$4\times 4$ matrix satisfying the two-mode symplectic condition   
\begin{eqnarray}\label{sympl}
R\Omega R^{\rm T}=\Omega,\quad \Omega=J\oplus J.
\end{eqnarray}
The two-mode symplectic transformations involve the unbalanced BS and QND
interaction (\ref{HQND}) as particular instances.
After that Alice completes the Bell measurement by homodyne detection of the position quadrature 
$x_{\rm in}'$ on mode ``in'' and the momentum quadrature $p_{A}''$ on mode $A$. 
Introducing the vector of the input quadratures $\xi_{\rm in}=(x_{\rm in},p_{\rm in})^{\rm T}$ the 
vector of the detected quadratures $\xi_{\rm d}=(x_{\rm in}',p_{A}'')^{\rm T}$ can be 
expressed as 
\begin{equation}\label{detected}
\xi_{\rm d}=Y\xi_{\rm in}+Z\xi_{A}',
\end{equation}
where
\begin{eqnarray}\label{XY}
Y=\left(\begin{array}{cc}
R_{33} & R_{34} \\
R_{23} & R_{24} \\
\end{array} \right),\quad
Z=\left(\begin{array}{cc}
R_{31} & R_{32} \\
R_{21} & R_{22} \\
\end{array} \right),
\end{eqnarray}
where $R_{kl}$ $(k,l=1,\ldots,4)$ are elements of the matrix $R$.
Alice measures certain classical values $\bar x_{\rm in}$ and $\bar p_{A}$ and sends them via 
classical channel to Bob who completes the teleportation by performing the displacements
\begin{equation}\label{displacement}
\xi_{B}'\rightarrow\xi_{\rm out}=\xi_{B}'+{\cal G}\bar\xi_{\rm d},
\end{equation}
where $\bar{\xi}_{\rm d}=(\bar{x}_{\rm in},\bar{p}_{A})^{\rm T}$ is
the c-number vector of the measurement results and  
\begin{eqnarray}\label{Gmatrix}
{\cal G}=\left(\begin{array}{cc}
{\cal G}_{xx} & {\cal G}_{xp} \\
{\cal G}_{px} & {\cal G}_{pp} \\
\end{array} \right)
\end{eqnarray}
is the matrix of the unnormalized gains. Here we restrict ourselves to the
unity gain regime in which the first moments of the input state are
preserved ($\langle\xi_{\rm out}\rangle=\langle\xi_{\rm in}\rangle$) and
therefore we chose
\begin{eqnarray}\label{G}
{\cal G}=Y^{-1}=\frac{1}{R_{24}R_{33}-R_{23}R_{34}}
\left(\begin{array}{cc}
R_{24} & -R_{34} \\
-R_{23} & R_{33} \\
\end{array} \right),
\end{eqnarray}
where $Y^{-1}$ is the inverse matrix to the matrix $Y$ that is assumed to be regular.
Hence, using Eqs.~(\ref{SA}), (\ref{detected}) and (\ref{displacement}) one finally finds
Bob's output quadrature operators in the form
\begin{eqnarray}\label{xpout}
\xi_{\rm out}=\xi_{\rm in}+Y^{-1}ZS_{A}\xi_{A}+S_{B}\xi_{B}.
\end{eqnarray}
A more instructive shape can be given to the latter formula. Defining the matrix 
$\Sigma\equiv \sigma_{3}Y^{-1}Z$, where $\sigma_{3}$ is the Pauli diagonal matrix 
$\sigma_{3}\equiv \mbox{diag}(1,-1)$, and taking into account the condition (\ref{sympl}) 
it can be shown that the new matrix $\Sigma$ satisfies the condition (\ref{singlesympl}) and it 
is therefore a symplectic matrix. Then the formula (\ref{xpout}) can be rewritten in the 
following form:
\begin{eqnarray}\label{finalxpout}
x_{\rm out}=x_{\rm in}+X,\quad p_{\rm out}=p_{\rm in}-P,  
\end{eqnarray}
where 
\begin{eqnarray}\label{genXP}
\left(\begin{array}{c}
X \\
P \\
\end{array}\right)=
\left(\begin{array}{c}
x_{A}''+x_{B}' \\
p_{A}''-p_{B}' \\
\end{array}\right)=\tilde S_{A}\xi_{A}+\sigma_{3}S_{B}\xi_{B} 
\end{eqnarray}
is the column vector of the quadrature added noises $X$ and $P$ and
$\tilde S_{A}=\Sigma S_{A}$ in another symplectic transformation.

The above analysis reveals that the use of a generic quadratic 
interaction $H_{R}$ in the Bell measurement has two effects. First, in order our 
teleportation protocol to preserve the first moments of the input state Bob musts 
the received classical outcomes of the Bell measurement 
$\bar{\xi}_{\rm d}=(\bar{x}_{\rm in},\bar{p}_{A})^{\rm T}$ 
transform by the matrix of gains ${\cal G}=Y^{-1}$ before using them 
for the displacements of his mode. For example, if the interaction $H_{R}$ is the 
QND interaction $H_{\rm QND}'=\kappa' x_{A}'p_{\rm in}$ considered in the 
previous section then ${\cal G}$ is the diagonal matrix ${\cal G}=\mbox{diag}(1,-1/g')$ 
($g'=\kappa't'$). Second, the interaction $H_{R}$ performs effectively a 
symplectic transformation $\Sigma$ on Alice's mode $A$. Now, consider the 
case when there are no auxiliary symplectic transformations on Alice's and Bob's sides 
($S_{A}=S_{B}=1$) and the state shared by Alice and Bob has the EPR
entanglement. In this case, the transformation $\Sigma$ entails that
the actual added noises $X$ and $P$ do not match with the ``optimal'' added noises 
$X_{\rm EPR}=x_{A}+x_{B}$, $P_{\rm EPR}=p_{A}-p_{B}$ and hence the amount of the noise 
added into the input state is larger in comparison with the ``optimal'' case. 
This undesirable effect of the transformation $\Sigma$ can be compensated by using  
an auxiliary symplectic transformation $S_{A}$ on mode $A$ of the form 
$S_{A,{\rm EPR}}=\Sigma^{-1}$. For instance, if the Bell measurement is realized by the QND 
interaction $H_{\rm QND}'$ (see Sec.~\ref{sec_QND}), then $\Sigma$ is the squeezing 
transformation $\Sigma_{\rm QND}=\mbox{diag}(g',1/g')$ and therefore 
$S_{A,{\rm EPR}}=\Sigma_{\rm QND}^{-1}=\mbox{diag}(1/g',g')$. 
If, on the other hand, Alice and Bob share a state with other than EPR entanglement and the 
interaction constant of the interaction in the Bell measurement can be controlled it 
can be adjusted in such a way that the effective transformation $\Sigma$ matches (on Alice's side)
the actual added noises with the ``optimal'' added noises and thus
enhances the teleportation fidelity. In the case of the entanglement produced by the 
QND interaction (\ref{HQND}) of two vacuum states the matching is achieved if 
$g'=(1+g^2)^{\frac{1}{4}}$ (see Sec.~\ref{sec_QND}).

Before going further, let us return for a while to the regularity requirement of the matrix $Y$ given in Eq.~(\ref{XY})
that emerged naturally when expressing the vector $\xi_{\rm in}$ from Eq.~(\ref{detected}).
Obviously, this property guarantees that the measured quadratures $x_{\rm in}'$ and $p_{A}''$
carry information about both complementary quadratures of the input state and therefore 
the first moments $\langle x_{\rm in}\rangle$ and $\langle p_{\rm in}\rangle$ of the input state 
can be preserved as is a natural demand for any unity gain teleportation protocol. 
Consequently, all the interactions yielding regular matrix $Y$ can be used in our scheme 
for the Bell measurement. For that reason we further restrict our attention to the interactions 
for which $Y$ is a regular matrix.

The quality of the CV teleportation protocols in a unity gain regime is well expressed 
quantitatively by the fidelity which is for pure input states defined as \cite{Braunstein_00}
\begin{eqnarray}\label{fidelitydef}
F=2\pi\int_{-\infty}^{+\infty}W_{\rm in}(x,p)W_{\rm out}(x,p)dxdp,
\end{eqnarray}
where $W_{\rm in}(x,p)$ and $W_{\rm out}(x,p)$ are Wigner functions of the input and output 
state, respectively. Assuming the input state, similarly as the shared state $\rho_{AB}$, 
to be described by a Gaussian Wigner function, expressing both the Wigner functions $W_{\rm in}$ and $W_{\rm out}$ 
as Fourier transforms of corresponding characteristic functions, carrying out all integrals and taking 
into account the relation $\langle\xi_{\rm out}\rangle=\langle\xi_{\rm in}\rangle$ following 
from the Eq.~(\ref{finalxpout}) we find that  
\begin{eqnarray}\label{fidelity}
F=\frac{1}{\sqrt{\mbox{det}\left(V_{\rm out}+V_{\rm in}\right)}},
\end{eqnarray}
where 
\begin{eqnarray}\label{varmatrix}
V_{i}=
\left(\begin{array}{cc}
\langle(\Delta x_{i})^{2}\rangle & \langle\{\Delta x_{i},\Delta p_{i}\}\rangle \\
\langle\{\Delta p_{i},\Delta x_{i}\}\rangle &\langle(\Delta p_{i})^{2}\rangle  \\
\end{array} \right),
\end{eqnarray}
is the single-mode variance matrix of the state $i$, $i={\rm out},{\rm in}$. 
Utilizing the formula (\ref{finalxpout}) one finds that the output variance matrix $V_{\rm out}$ 
is of the form:
\begin{equation}\label{Vout}
V_{\rm out}=V_{\rm in}+2N,
\end{equation}
where 
\begin{eqnarray}\label{N}
2N=\left(\begin{array}{cc}
\langle X^{2}\rangle & -\langle\{X,P\}\rangle \\
-\langle\{P,X\}\rangle &\langle P^{2}\rangle  \\
\end{array} \right).
\end{eqnarray}
Substituting the last expression of the Eq.~(\ref{genXP}) for the added noises $X$ and $P$
into the matrix (\ref{N}) it can be expressed in the following matrix form:  
\begin{eqnarray}\label{Nmatrix}
2N&=&\sigma_{3}\tilde S_{A}A\tilde S_{A}^{\rm T}\sigma_{3}+S_{B}BS_{B}^{\rm T}
+\sigma_{3}\tilde S_{A}CS_{B}^{\rm T}\nonumber\\
&&+S_{B}{C}^{\rm T}\tilde S_{A}^{\rm T}\sigma_{3},
\end{eqnarray}
where $A$, $B$, $C$ and $C^{\rm T}$ are $2\times 2$ blocks of the variance matrix $V_{AB}$ of 
the shared state, i. e.  
\begin{eqnarray}\label{cormatrix}
V_{AB}=\left(\begin{array}{cc}
A & C \\
C^{\rm T} & B \\
\end{array} \right).
\end{eqnarray}
The fidelity (\ref{fidelity}) attains particularly simple form in the
case of teleportation of coherent states when $V_{\rm in}=(1/2)I$. Combining the
Eqs.~(\ref{fidelity}) and (\ref{Vout}) the fidelity then reads as 
\begin{eqnarray}\label{fidelityfinal}
F_{\rm coh}=\frac{1}{\sqrt{1+2\mbox{Tr}N+4\mbox{det}N}}.
\end{eqnarray}
The quality of our teleportation is determined by the function $2\mbox{Tr}N+4\mbox{det}N$. 
For a fixed shared entangled state and a fixed interaction in the Bell measurement this is a 
complicated function of six real parameters of the symplectic transformations $S_{A}$ 
and $S_{B}$. Hence, the maximization of the fidelity (\ref{fidelityfinal}) with respect to  
the transformations $S_{A}$ and $S_{B}$ amounts to finding the minimum of the function of 
several real variables which is a task that can be hardly solved even
numerically. Instead of doing this we will minimize another more simple measure of
teleportation success given by the noise ${\cal N}$ defined through the formula \cite{Horoshko_00}
\begin{eqnarray}\label{mean}
\langle n_{\rm out}\rangle=\langle n_{\rm in}\rangle+{\cal N},
\end{eqnarray}
where $\langle n_{\rm out}\rangle$ and $\langle n_{\rm in}\rangle$ are  
the mean numbers of photons in the output and input state, respectively 
\cite{Horoshko_00}. Calculating $\langle n_{\rm out}\rangle$ using 
the Eq.~(\ref{Vout}) one finds that the noise ${\cal N}$ is
connected with the matrix (\ref{N}) by the formula  
\begin{eqnarray}\label{N'}
{\cal N}=\mbox{Tr}N=\left[\langle(x_{A}''+x_{B}')^{2}\rangle+\langle (p_{A}''-p_{B}')^{2}\rangle\right]/2.
\end{eqnarray}
To make the problem more tractable we consider that the shared state is a 
fixed pure entangled two-mode Gaussian state. In this case there exist local 
symplectic transformations $m_{A}$ and $m_{B}$ that bring the   
variance matrix (\ref{cormatrix}) to the so called standard form \cite{Vidal_02}
$V_{\rm TMS}=\left(m_{A}\oplus m_{B}\right)V_{AB}\left(m_{A}\oplus m_{B}\right)^{\rm T}$, 
where 
\begin{eqnarray}\label{standard}
V_{\rm TMS}=\left(\begin{array}{cccc}
a & 0 & -c & 0 \\
0 & a & 0 & c \\
-c & 0 & a & 0 \\
0 & c & 0 & a\\
\end{array}\right),
\end{eqnarray}
where $a=\sqrt{\mbox{det}A}\geq1/2$, $c=\sqrt{|\mbox{det}C|}>0$ and $a^{2}-c^{2}=1/4$. 
Owing to the last equality and the inequality $c>0$ we can put $a=\cosh(2\kappa)/2$ and $c=\sinh(2\kappa)/2$, where 
$\kappa>0$ and thus the variance matrix (\ref{standard}) describes the 
two-mode squeezed (TMS) vacuum state having squeezed variances 
$\langle(x_{A}+x_{B})^{2}\rangle_{\rm TMS}=\langle(p_{A}-p_{B})^{2}\rangle_{\rm TMS}=
2(\sqrt{\mbox{det}A}-\sqrt{|\mbox{det}C|})=e^{-2\kappa}$ 
($\kappa$ is the squeezing parameter and $\langle\,\,\,\rangle_{\rm TMS}$ denotes averaging over the TMS
vacuum state (\ref{standard})) and therefore possessing the EPR entanglement. 
Expressing now the symplectic matrices $\tilde S_{A}$ and $S_{B}$ in the Eq.~(\ref{Nmatrix}) in the form 
$\tilde S_{A}=s_{A}m_{A}$ and $S_{B}=s_{B}m_{B}$, where $s_{A}$, $s_{B}$ are some new symplectic 
matrices, utilizing the Eq.~(\ref{standard}) and substituting the obtained formula for the matrix $N$ to
the the definition (\ref{N'}) one finds that 
\begin{eqnarray}\label{Nadd}
{\cal N}&=&\frac{a}{2}\mbox{Tr}(\tilde s_{A}\tilde s_{A}^{\rm T}+
s_{B}s_{B}^{\rm T})-c\mbox{Tr}(\tilde s_{A}s_{B}^{\rm T}),
\end{eqnarray}
where $\tilde s_{A}=\sigma_{3}s_{A}\sigma_{3}$. Further, according to the Bloch-Messiah reduction 
\cite{Braunstein_99} one can decompose the symplectic matrices $\tilde s_{A}$ and $s_{B}$ as 
follows: 
\begin{equation}\label{BM}
\tilde s_{A}=P(\alpha)S(r_{A})P(\beta),\quad s_{B}=P(\gamma)S(r_{B})P(\delta),
\end{equation}
where
\begin{eqnarray}\label{RS}
P(u)=\left(\begin{array}{cc}
\cos u & -\sin u \\
\sin u & \cos u \\
\end{array} \right),\quad
S(v)=\left(\begin{array}{cc}
e^{v} & 0 \\
0 & e^{-v} \\
\end{array} \right)
\end{eqnarray}
are the matrix of the phase shift by $u=\alpha, \beta, \gamma, \delta$ and the matrix of 
the single-mode squeezer with the squeezing parameter $v=r_{A}, r_{B}$, respectively. On inserting
expressions (\ref{BM}) and (\ref{RS}) into the Eq.~(\ref{Nadd}) and using the invariance of the trace 
with respect to cyclic permutation of its arguments we arrive at the formula
\begin{eqnarray}\label{Np}
{\cal N}&=&2a\cosh r_{+}\cosh r_{-}-c\left[(\cosh r_{+}+\cosh r_{-})\cos\theta_{+}\right.
\nonumber\\
&&\left.+(\cosh r_{+}-\cosh r_{-})\cos\theta_{-}\right],
\end{eqnarray}
where $r_{\pm}=r_{A}\pm r_{B}$ and $\theta_{\pm}=\alpha\pm\beta-\gamma\mp\delta$. Clearly, 
${\cal N}$ is minimized with respect to $\theta_{+}$ when $\theta_{+}=2k\pi$, where $k$ is an integer. 
Minimization with respect to $\theta_{-}$ requires to distinguish three cases: (i) 
$\cosh r_{+}>\cosh r_{-}$, (ii) $\cosh r_{+}<\cosh r_{-}$, and (iii) $\cosh r_{+}=\cosh r_{-}$.
The values of $\theta_{-}$ minimizing ${\cal N}$ are equal to $\theta_{-}=2l\pi$ in case (i),  
$\theta_{-}=(2l+1)\pi$ in case (ii) ($l$ is an integer in both the cases), and $\theta_{-}$ can be 
arbitrary in the last case (iii). In addition, in all the three cases the necessary 
conditions on the extreme, $\partial{\cal N}/\partial r_{\pm}=0$ are satisfied when 
$r_{\pm}=0$. Since $\cosh r_{+}=\cosh r_{-}$ for $r_{+}=r_{-}=0$ the function (\ref{Np}) can 
have extreme only on the boundary characterized by the condition (iii), where it reads
\begin{eqnarray}\label{Nadd'}
{\cal N}'&=&2\cosh r_{+}(a\cosh r_{+}-c\cos\theta_{+}).
\end{eqnarray}
In order the candidates for extrema localized in points for which $\theta_{+}=2k\pi$ 
and $r_{\pm}=0$ to be minima (no further condition is obtained as $\theta_{-}$ can be arbitrary) 
the sufficient condition on the minimum must be satisfied. It is given by the positive 
definiteness of the matrix ${\cal A}$ with elements        
\begin{eqnarray}\label{second}
{\cal A}_{11}=\frac{\partial^{2}{\cal N}'}{\partial r_{+}^{2}},\enspace 
{\cal A}_{22}=\frac{\partial^{2}{\cal N}'}{\partial \theta_{+}^{2}},\enspace
{\cal A}_{12}={\cal A}_{21}=\frac{\partial^{2}{\cal N}'}{\partial r_{+}\partial \theta_{+}}.
\nonumber\\
\end{eqnarray}
Since ${\cal A}_{11}=2(2a-c)>0$, ${\cal A}_{22}=2c>0$ and ${\cal A}_{12}=0$ in the potential 
extrema as follows from the inequalities $a-c>0$ and $c>0$ the matrix ${\cal A}$ is positive 
definite and hence the potential extrema are minima all giving the same value
\begin{eqnarray}\label{Naddmin}
{\cal N}_{\rm min}&=&2(a-c)=2(\sqrt{\mbox{det}A}-\sqrt{|\mbox{det}C|})\nonumber\\ 
&=&\frac{\langle(x_{A}+x_{B})^{2}\rangle_{\rm TMS}+\langle (p_{A}-p_{B})^{2}\rangle_{\rm TMS}}{2}\nonumber\\
&=&e^{-2\kappa}.
\end{eqnarray}
The equalities $r_{\pm}=0$ and $\theta_{+}=2k\pi$ imply that 
$r_{A}=r_{B}=0$ and $\gamma+\delta=\alpha+\beta-2k\pi$. From the Eq.~(\ref{BM}) it then follows that the transformations 
minimizing (\ref{Nadd}) are the arbitrary same phase shifts $\tilde s_{A}=s_{B}=P(\alpha+\beta)$ 
where $\alpha+\beta$ is an arbitrary angle. This means that the noise ${\cal N}$ is 
minimum if the transformations $S_{A}$ and $S_{B}$ are such that  
$S_{A}=\Sigma^{-1}m_{A}$ and $S_{B}=m_{B}$ ($\alpha+\beta=0$ was chosen for simplicity) 
where $m_{A}$, $m_{B}$ bring the shared state (\ref{cormatrix}) to the TMS vacuum state 
(\ref{standard}). From the Eq.~(\ref{Nmatrix}) then follows that $2N={\cal N}_{\rm min}I$ 
and therefore the fidelity (\ref{fidelityfinal}) reads
\begin{eqnarray}\label{fpur}
F_{\rm coh}=\frac{1}{1+{\cal N}_{\rm min}}=\frac{1}{1+e^{-2\kappa}} 
\end{eqnarray}
and coincides with the fidelity of the BK scheme. Thus, we have shown that  
if we characterize our generalized CV teleportation scheme by the noise 
${\cal N}$ then its performance can be improved by the suitable auxiliary local symplectic 
transformations on parts of the shared entangled state. The optimum in which the noise ${\cal N}$ 
attains minimum is achieved for those local symplectic transformations that convert effectively 
our scheme to the BK scheme, i. e. the BK scheme is optimal from the point of view of the noise 
${\cal N}$. As a by-product we have also proved that on the set of pure
two-mode Gaussian states that can be produced from the TMS vacuum state (\ref{standard}) by 
{\it all} local symplectic transformations the total variance 
$\left[\langle(x_{A}+x_{B})^{2}\rangle+\langle (p_{A}-p_{B})^{2}\rangle\right]/2$ is
minimized by the TMS vacuum state. This proof generalizes for pure states the proof given 
recently in \cite{Adesso_04} in which the minimization of the total variance was performed 
only with respect to a two-parametric subset of all local symplectic transformations that is formed by
the symplectic transformations that can be expressed as a product of a squeezer and a phase shift.  
\section{Conclusion}\label{sec_conclusion}

In conclusion, we have investigated the CV teleportation protocol 
based on the shared entanglement produced by the QND interaction 
of two vacua and utilizing QND interaction or an unbalanced BS in 
the Bell measurement. We have shown on two particular examples 
that in the non-unity gain regime the signal transfer coefficient 
can be increased while the conditional variance product remains 
unchanged by a suitable squeezing operation on sender's side of 
the shared entangled state. Further, it was demonstrated that in 
the unity gain regime the teleportation fidelity can be enhanced 
if the shared entanglement is redistributed by local squeezing 
operations converting effectively the protocol to the standard BK 
teleportation protocol. Finally, it is proved that such a choice of 
the local squeezing operations is optimal if we characterize our 
teleportation scheme by the noise by which the mean number of photons in the 
input state is increased in the teleportation process.
\medskip
\section*{Acknowledgments}
We thank G. Leuchs and N. Korolkova for the kind hospitality at the Erlangen University. 
The contributions of P. van Loock and J. Fiur\'a\v{s}ek are also gratefully acknowledged. 
The research has been supported by Project LN00A015, Research Project No. CEZ: J14/98 of 
the Czech Ministry of Education and EU under project COVAQIAL (FP6-511004). 
R. F. acknowledges support by Project 202/03/D239 of the Grant Agency of the Czech Republic.

\end{document}